\documentclass[aps,prb,twocolumn,showpacs,superscriptaddress,floatfix]{revtex4-1}
\usepackage{amsfonts}
\usepackage{amsmath}
\usepackage{amssymb}
\usepackage{graphicx}
\usepackage{float}
\usepackage{mathtools}

\DeclareFontEncoding{LGR}{}{}
\DeclareTextSymbol{\~}{LGR}{126}
\newcommand{\lyxmathsym}[1]{\ifmmode\begingroup\def\b@ld{bold}
 \text{\ifx\math@version\b@ld\bfseries\fi#1}\endgroup\else#1\fi}

\begin{document}

\title{Influence of ageing on Raman spectra and the conductivity of monolayer
graphene samples irradiated by heavy and light ions}

\date{\today }

\author{A. Butenko}

\affiliation{Institute of Nanotechnology and Advanced Materials, Bar-Ilan University,
Ramat Gan 52900, Israel }

\author{E. Zion}

\affiliation{Institute of Nanotechnology and Advanced Materials, Bar-Ilan University,
Ramat Gan 52900, Israel }

\author{Yu. Kaganovskii}

\affiliation{Jack and Pearl Resnick Institute, Department of Physics, Bar-Ilan
University, Ramat Gan 52900, Israel }

\author{L. Wolfson}

\affiliation{Jack and Pearl Resnick Institute, Department of Physics, Bar-Ilan
University, Ramat Gan 52900, Israel }

\author{V. Richter}

\affiliation{Institute of Nanotechnology and Advanced Materials, Bar-Ilan University,
Ramat Gan 52900, Israel }

\author{A. Sharoni}

\affiliation{Institute of Nanotechnology and Advanced Materials, Bar-Ilan University,
Ramat Gan 52900, Israel }

\author{E. Kogan}

\affiliation{Jack and Pearl Resnick Institute, Department of Physics, Bar-Ilan
University, Ramat Gan 52900, Israel }

\author{M. Kaveh}

\affiliation{Jack and Pearl Resnick Institute, Department of Physics, Bar-Ilan
University, Ramat Gan 52900, Israel }

\author{I. Shlimak}

\affiliation{Jack and Pearl Resnick Institute, Department of Physics, Bar-Ilan
University, Ramat Gan 52900, Israel }
\begin{abstract}
The influence of long-term ageing (about one year) on the Raman scattering
(RS) spectra and the temperature dependence of conductivity has been
studied in two series of monolayer graphene samples irradiated by
different doses of C$^{+}$ and Xe$^{+}$ ions. It is shown that the
main result of ageing consists of changes in the intensity and position
of D- and G- and 2D-lines in RS spectra and in an increase of the
conductivity. The observed effects are explained in terms of an increase
of the radius of the \textquotedblleft activated\textquotedblright{}
area around structural defects.
\end{abstract}

\pacs{73.22.Pr, 72.80.Vp}

\maketitle

\section{introduction}

Ion irradiation is widely used for introducing defects and disorder
in graphene films in order to modify their physical properties (see,
for example, \cite{1,2,3,4}). The measurement of the Raman scattering (RS)
spectra is considered an effective tool for probing the structure
of disordered graphene and the density of introduced defects \cite{5,6,7,8}.
Typical RS spectra for disordered graphene consist of three main lines.
The D-line at 1350 cm$^{-1}$ is related to the inter-valley double
resonant process in graphene in the vicinity of a lattice defect (edge,
vacancies.) The 2D-line at 2700 cm$^{-1}$, is related to an inter-valley
two phonon mode, characteristic of the perfect crystalline honeycomb
structure. The G-line at 1600 cm$^{-1}$ is common for different carbon-based
materials, including carbon nanotubes and graphite. The appearance
of the D-peak is ascribed to the breathing of carbon hexagons at the
borders of the crystallite areas or in the vicinity of a structural
defect, reflecting the loss of translational symmetry \cite{5}. The
intensity of the D-line is used (usually in the form of the dimensionless
ratio of intensities of the D- and G-lines, $I_{D}/I_{G}$) as a measure
of disorder in the graphene layer. In addition, a few minor lines
connected with different modes of phonons can be seen in disordered
samples: D'-line (1620 cm$^{-1}$), D+D'-line (2970 cm$^{-1}$) and
D+D''-line (2450 cm$^{-1}$). The positions of all lines are given
for 532 nm excitation laser.\par
The main aim of the present study is
the investigation of the influence of long-term ageing (about one
year) on properties of monolayer graphene samples irradiated by different
dose of heavy and light ions. From one side, the ambient atmosphere is not fully reproducible in different places and laboratories, but from the other side, most devices are kept in ambient atmosphere, therefore this investigation might be of interest for specialists in graphene-based devices.

\section{samples}

In our previous studies \cite{9,10,11}, we reported the results of the
investigation of the optical (RS) and the electrical (conductivity
and magnetoresistance) properties of monolayer graphene samples irradiated
by different doses $\Phi$ of carbon ions C$^{+}$ with energy
35 keV. Six series of micro-samples (200x200 $\mu$m) provided with two metal contacts were prepared by means of electron-beam lithography on a common large-scale (5x5 mm) monolayer graphene
specimen supplied by \textquotedblleft Graphenea\textquotedblright{}
company. Monolayer graphene was produced by CVD on copper catalyst
and transferred to a 300 nm SiO$_{2}$/Si substrate using wet transfer
process. Graphene film of such a large size was not a monocrystalline. It looks like a polycrystalline film with the average size of microcrystals about a few microns \cite{9}. In the present study, we also include the results obtained
on a series of monolayer graphene samples irradiated by heavy ions
Xe$^{+}$ with the same energy $E$ = 35 keV. In this series, micro-samples were not fabricated, six areas 2x1 mm of the whole specimen 5x5 mm were irradiated separately with different doses of Xe ions.

\section{raman scattering}

In the RS spectra measurements, excitation was realized by the laser beam with $\lambda$ = 532 nm and power less than 2 mW to avoid heating and film destruction. Reproducibility was verified by repeated measurements in different areas and different samples of the same group.\\

\begin{figure}[H]
	\begin{centering}
		\includegraphics[scale=0.8]{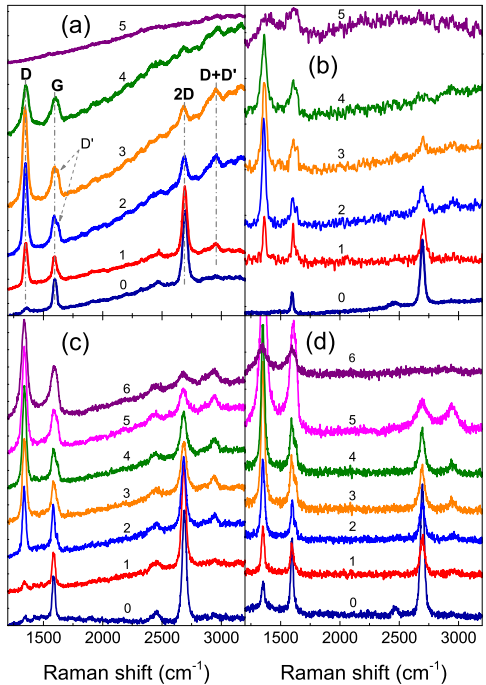}
		\par\end{centering}
	
	\caption{(Color online) Raman scattering spectra for two series of samples irradiated with
		C$^{+}$ ions (a, b) and Xe$^{+}$ ions (c, d): (a, c) - after irradiation,
		(b, d) - after long-term ageing. For C-series, $\Phi$ in units
		of 10$^{14}$ cm$^{-2}$: 0 \textendash{} 0 (initial), 1 \textendash{} 0
		(non-irradiated, but after E-beam lithography {[}9{]}), 2 \textendash{}
		0.5, 3 \textendash{} 1.0, 4 \textendash{} 2.0, 5 \textendash{} 10;
		for Xe-series, $\Phi$ in units of 10$^{13}$ cm$^{-2}$: 0 \textendash{}
		0 (initial), 1 \textendash{} 0.15, 2 \textendash{} 0.3, 3 \textendash{}
		0.5, 4 \textendash{} 1.0, 5 \textendash{} 2.0, 6 \textendash{} 4.0.}
	\label{Layout1}
\end{figure}

\begin{figure}[H]
\hspace*{1cm}\includegraphics[scale=0.8]{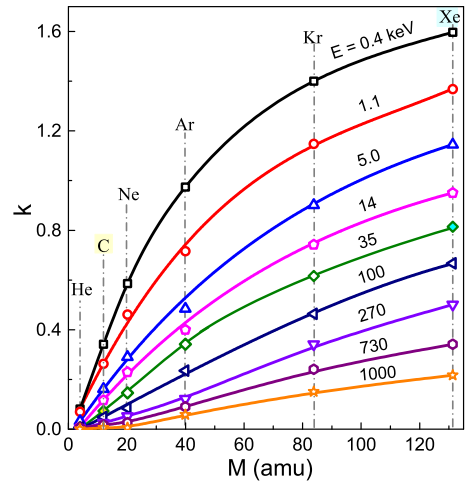}

\caption{(Color online) Dependence of $k$ on the ion mass $M$ for different energies $E$
shown in keV. Masses of some ions are indicated by vertical dotted
lines.}
\label{Layout2}
\end{figure}

Figure~$\ref{Layout1}$ shows the RS spectra for both series of samples
immediately following irradiation (a, c) and after long-term ageing
(b, d). The changes of the main lines of the RS with increasing $\Phi$
are as expected: in non-irradiated samples (0) the intensity of the
\textquotedblleft defective\textquotedblright{} D-line is small or
absent which demonstrates the good quality of the initial monolayer
graphene specimen. Increasing the dose of the irradiation $\Phi$
causes the D-line to increase and then broaden and decrease, while
the 2D-line monotonically decreases and disappears. To compare results
of irradiation by different ions, one plots the ratio $I_{D}/I_{G}$
and $I_{2D}/I_{G}$ as a function of the density of structural defects
$N_{D}$, which is proportional to the dose of irradiation $\Phi$,
$N_{D}=k$$\Phi$. The coefficient $k$ depends on the energy
$E$ and the mass $M$ of the irradiated ions and reflects the average
fraction of carbon vacancies in the graphene lattice per ion impact.
A computer simulation of this process was performed in Ref. \cite{12}.
On the basis of this simulation, we plot in Fig.~$\ref{Layout2}$
the dependence of $k$ as a function of $M$ (in atomic mass units,
amu) for different $E$.

\begin{figure}[H]
\begin{centering}
\includegraphics[scale=0.8]{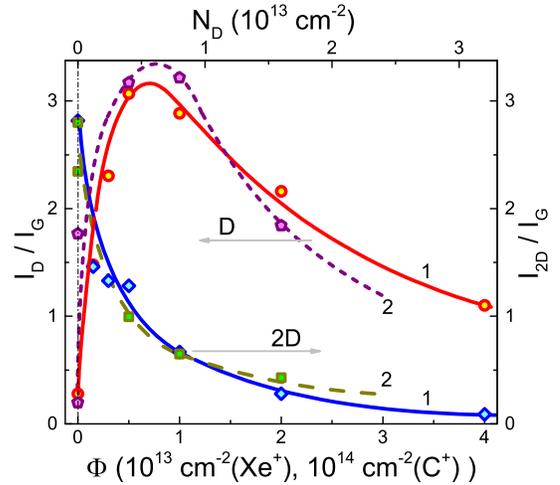}
\par\end{centering}

\caption{(Color online) The ratio $I_{D}/I_{G}$ and $I_{2D}/I_{G}$ for \textquotedblleft irradiated\textquotedblright{}
samples for Xe- (1) and C- (2) series as a function of $\Phi$
(in units of 10$^{13}$ cm$^{-2}$ for Xe$^{+}$ and 10$^{14}$ cm$^{-2}$
for C$^{+}$). The upper scale shows the density of introduced defects
$N_{D}$ common for both series.}
\label{Layout3}
\end{figure}

One can see in Fig.~$\ref{Layout2}$ from the curve $E$ = 35 keV
that for C$^{+}$ ions ($M$ = 12) $k$ \ensuremath{\approx} 0.06-0.08,
whereas for Xe$^{+}$ ($M$ = 131), $k$ \ensuremath{\approx} 0.8.
Therefore, to equalize the density of the introduced defects $N_{D}$,
we used for Xe$^{+}$ one order of magnitude smaller doses $\Phi$
than for C$^{+}$ ions. The dependence of $I_{D}/I_{G}$ and $I_{2D}/I_{G}$
for both series of samples after irradiation is shown in Fig.~$\ref{Layout3}$.
One can see that the experimental data for the C- and Xe- series are
indeed close but not identical.

\begin{figure}[H]
\begin{centering}
\includegraphics[scale=0.7]{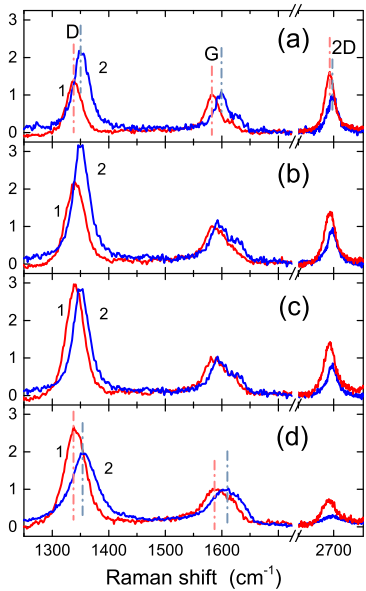}
\par\end{centering}

\caption{(Color online) Comparison of the normalized RS spectra in \textquotedblleft just irradiated\textquotedblright{}
(1) and \textquotedblleft aged\textquotedblright{} (2) samples from
Xe-series with different irradiation dose $\Phi$ (in units
of 10$^{13}$ cm$^{-2}$): $a-0.3,b\lyxmathsym{\textendash}0.5,c-1.0,d\lyxmathsym{\textendash}2.0$.
Intensity of G-line is taken as 1.}
\label{Layout4}
\end{figure}

Long-term ageing leads to changes in the RS spectra. Figure~$\ref{Layout4}$
shows the spectra for some ``irradiated'' and ``aged'' samples
from the Xe-series. One can see that the intensity of D-lines in the
``aged'' samples is stronger for weakly irradiated samples but weaker
for strongly irradiated samples, while the 2D-lines are always weaker
in the ``aged'' samples.

\begin{figure}[H]
\begin{centering}
\includegraphics[scale=0.7]{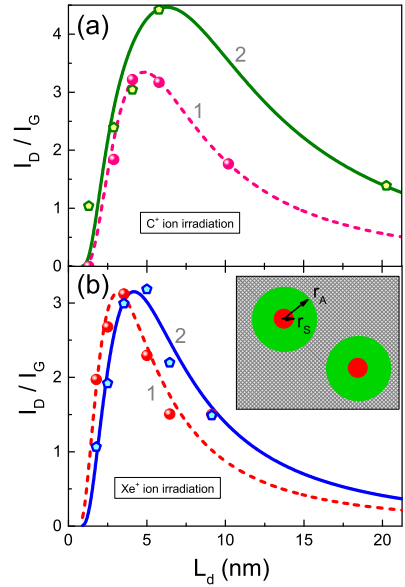}
\par\end{centering}

\caption{(Color online) $I_{D}/I_{G}$ as a function of $L_{D}$ for C-series (a) and Xe-series
(b) of samples. Curves 1 and 2 show the best fit of theoretical curve
Eq. (\ref{Eq1}) for ``irradiated'' and ``aged'' samples, correspondingly.
Insert shows schematics of radiation defects of two length scales
(see text).}
\label{Layout5}
\end{figure}

To explain the changes in the RS induced by long-term ageing, we plot
in Figs.~$\ref{Layout5}$a,b the ratio $I_{D}/I_{G}$ as a function of the mean distance
between defects, $L_{D}\approx N_{D}^{-1/2}$. At large $L_{D}$ (small
$\Phi$), the ratio $I_{D}/I_{G}$ increases with decreasing
$L_{D}$, passes through a maximum and then rapidly decreases. Such
non-monotonic behavior is common for many experimental observations
\cite{5,6,7,8,9} and agrees with the theoretical model \cite{13} based on
the assumption that a single ion impact leads to the formation of
a defect characterized by two length scales $r_S$ and $r_A$ ($r_A>r_S$),
which are the radii of two circular areas surrounding the defect (see
insert in Fig.~$\ref{Layout5}$). The area $S$ within the shorter
radius $r_S$ around the impact point is structurally disordered. In
the $A$-area (\textquotedblleft activated\textquotedblright{} area)
between $r_A$ and $r_S$, the lattice structure is preserved, but the
proximity to the impact point causes a breakdown of the selection
rules and gives rise to the \textquotedblleft defective\textquotedblright{}
D-peak. The 2D-peak, which is characteristic for a perfect graphene
lattice, can be generated only if the electron-hole excitation occurs
outside the A-area. Increasing the irradiation dose $\Phi$
obviously results in an increase of the D-line intensity and a decrease
of the 2D-line intensity. Increasing of $\Phi$ leads to a decrease
of $L_{D}\approx(k\Phi)^{-1/2}$ and when $L_{D}$ becomes shorter
than $r_A$, the $A$-areas begin to overlap with each other and with
$S$-areas. As a result, $I_{D}/I_{G}$ passes through a maximum and
then decreases.

The solid and dotted lines in Fig.~$\ref{Layout5}$ represent the
theoretical expression for the dependence of $I_{D}/I_{G}(L_{D})$
that was obtained in Ref. \cite{13} and modified in Ref. \cite{9}:

\begin{equation} \label{Eq1}
\begin{split}
I_D/I_G &= C_Ae^{-\pi r_S^{2}/L_D^{2}}[1-e^{-\pi(r_A^2-r_S^2)/L_D^{2}}] \\
& +C_S[1-e^{-\pi r_S^{2}/L_D^{2}}]
\end{split}
\end{equation}

Here, the parameter $C_{A}$ is the maximal possible value of the
$I_{D}/I_{G}$ ratio in graphene, which could be attained for the
situation in which no damage occurred to the hexagonal network of
carbon atoms, and $C_{S}$ is the $I_{D}/I_{G}$ value in the highly
disordered limit \cite{5}.

The best fit of the experimental data shown in Fig.~$\ref{Layout5}$
is given by the solid and dashed lines, corresponding to the following
values of the adjustable parameters in Eq. (\ref{Eq1}):

\begin{table}[H]
\centering{}%
\begin{tabular}{|c|c|}
\hline
C-series & Xe-series\tabularnewline
\hline
$C_{S}=0,C_{A}=5.4,$ & $C_{S}=0,C_{A}=4.9,$\tabularnewline
\hline

\begin{tabular}{c|c}
``irradiated'' & ``aged''\tabularnewline
\hline
$r_{S}$(nm): 1.55 & 1.35\tabularnewline
\hline
$r_{A}$(nm): 4.1 & 6.4\tabularnewline
\end{tabular} & %
\begin{tabular}{c|c}
``irradiated'' & ``aged''\tabularnewline
\hline
 1.0 & 1.3\tabularnewline
\hline
 2.8 & 3.7\tabularnewline
\end{tabular}\tabularnewline
\hline
\end{tabular}
\caption{Adjustable parameters of Eq.(1) determined from the best fitting with experimental data shown in Fig. 5.}
\end{table}

Analysis of these data shows that the main effect occurs as a function
of $r_{A}$. Comparison of ``irradiated'' samples before ageing
shows that irradiation with relatively light ions (C$^{+}$) is characterized
by a larger values of $r_{S}$ and $r_{A}$ than irradiation with
heavy ions (Xe$^{+}$). This could be connected with existence of
a backscattering in the case of bombardment of SiO$_{2}$ substrate
by C ions which are lighter than atoms in the substrate, in contrast
to irradiation by heavy Xe ions.

Long-term ageing leads to an increase in $r_{A}$ in both series.
The increase of the square of $A$-area could be caused by the capture
of molecules from the ambient atmosphere which can lead to increasing
the Raman relaxation length $l=r_{A}\lyxmathsym{\textendash}r_{S}$.
\cite{5} This increase of $r_{A}$ explains the different behavior
of the D-line for weakly and strongly irradiated samples shown in
Fig.~$\ref{Layout4}$. For weakly irradiated samples, the D-line
increases because of the increase of the square of $A$-areas. However,
the overlap of increased $A$-areas begins earlier, at a lower density
of defects, which explains the decrease of the D-line in strongly
irradiated samples.

One can see also from Fig.~$\ref{Layout4}$ that the peaks in the
\textquotedblleft aged\textquotedblright{} samples shift to higher
frequencies (\textquotedblleft blue shift\textquotedblright ). The
largest values of the blue shift are 18 and 14 cm$^{-1}$ for the
G- and D-peaks, respectively. There are several possible origins for
the blue shift, including stress and the effects of doping \cite{14}.
The doping-based scenario agrees with the assumption that the increase
of $r_{A}$ in aged samples is caused by the capture of molecules
from the ambient atmosphere.

\section{electrical conductivity}

\begin{figure}[H]
\hspace*{0.5cm}\includegraphics[scale=0.7]{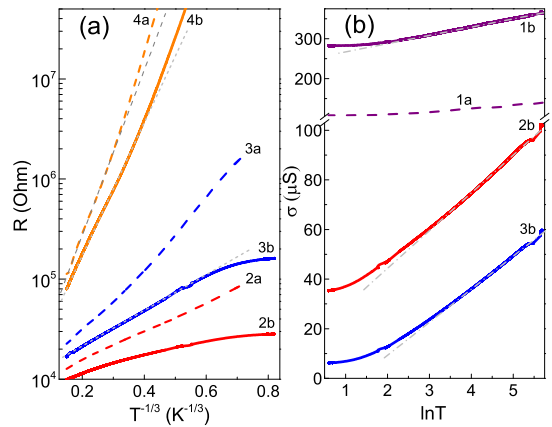}
\caption{(Color online) Temperature dependence of resistivity $R$ and conductivity $\sigma=1/R$
of C$^{+}$-irradiated samples plotted on the scale appropriate for
VRH regime (a) and for WL regime (b). Numbers near curves correspond
to the numbers of samples in Fig.~$\ref{Layout1}$(a,c): 1a, 2a, 3a, 4a \textendash{}
\textquotedblleft irradiated\textquotedblright{} samples, 1b, 2b,
3b, 4b \textendash ''aged'' samples.}
\label{Layout6}
\end{figure}

In series of C$^{+}$-irradiated samples,
the temperature dependence of the conductivity $\sigma$ and the resistivity
$R=1/\sigma$ were measured before ageing, after irradiation, and
the results were reported in Refs. \cite{10,11}. It was shown that
$\sigma(T)$ for sample 1 with a low density of defects, $N_{D}$,
is accurately described in the framework of the weak localization
(WL) model, whereby $\sigma(T)\sim ln(T)$ and saturates at low $T$. The resistivity of more disordered samples 2,3,4 is governed
by variable-range hopping (VRH), characteristic of strongly localized
carriers. VRH is described by exponential laws: $R\sim exp(T^{-p})$,
with $p=1/3$ at relatively high $T$ (``Mott-law'') and $p=1/2$
(``Efros-Shklovski law'') at low $T$. Measurements of $R(T)$
after ageing showed the increase of conductivity for all samples (Fig.~$\ref{Layout6}$).\\

On the first glance, increase of conductivity could be attributed to the doping due to capture of molecules from the ambient atmosphere. However, the term \textquotedblleft doping\textquotedblright is more applicable to pristine or slightly disordered graphene when conductivity is realized by free charge carriers and can be changed by variation of the gate voltage in the field-effect-transistor (FET) geometry in both directions from the charge neutrality point \cite{15,16}. in our samples, variation of the gate voltage up to +/- 80 V weakly changes the conductivity, so the charge neutrality point was not achieved. This is due to the fact, that conductivity in our disordered highly resistive graphene samples is realized by localized charge carriers \cite{10,11}. Localization is caused by the random potential relief, induced by the large density of negatively and positively charged defects. Increase of $r_{A}$ can lead to smearing of this relief, weakening of localization and corresponding increase of conductivity. Therefore an increase of  $r_{A}$ due to capture of molecules from the ambient atmosphere would be better described as \textquotedblleft alteration\textquotedblright rather than \textquotedblleft doping\textquotedblright. \\

Figure~$\ref{Layout6}$ shows that before ageing, $R(T)$ for \textquotedblleft irradiated\textquotedblright
samples 2 and 3 is governed by the VRH mechanism which corresponds
to a straight line when plotted as $log(R)$ vs. $T^{-1/3}$ (curves
2a and 3a in Fig.~$\ref{Layout6}$a), while after ageing, the reduced
resistance deviates from a straight line and even shows a tendency
to saturation at low $T$ (curves 2b and 3b in Fig.~$\ref{Layout6}$a).
As a result, the conductivity of these samples is better described
by a plot of WL $\sigma(T)\sim ln(T)$ (Fig.~$\ref{Layout6}$b). The
transition from strong to weak localization after ageing can be explained
by a softening of the random potential relief due to an increase of
$r_{A}$. For strongly irradiated sample 4, the dependence $R(T)$
after ageing still exhibits the VRH mechanism even though the resistivity
decreases.\\

The influence of ageing on the Raman scattering spectra and the conductivity in graphene layers had not been studied earlier. There were only measurements of the Raman signal from aged nanographite \cite{17} and from organic molecules absorbed on the aged graphene \cite{18}. In Ref.\cite{16}, an increase of resistance after ageing in ambient atmosphere was observed in field effect transistors made from pristine graphene. In Ref. \cite{19}, the influence of ageing was studied on the electrical properties of graphene-based transparent flexible electrodes.\\

In this study, we
show that long-term ageing of pristine and ion-irradiated monolayer
graphene samples kept in ambient atmosphere leads to changes in the
intensity and a blue-shift in the position of the D-, G- and 2D-lines
in the RS spectra and to an increase of the electrical conductivity.
Our explanation is based on the notion of \textquotedblleft activated\textquotedblright{}
areas around structural defects in graphene. We suggest that the radius
of this ``activated'' areas increases in the process of ageing which
is caused by capturing molecules from the ambient atmosphere.

\section{acknowledgements}

One of the authors (E.K.) cordially thanks Center for Theoretical
Physics of Complex Systems, Institute for Basic Science (IBS), Daejeon,
Republic of Korea for the hospitality during his stay.

\end{document}